\documentclass[12pt]{iopart}

\usepackage{amsmath}
\usepackage{amsfonts}
\usepackage{amssymb}
\usepackage{amsthm}
\usepackage{iopams}
\usepackage[cp1251]{inputenc}
\usepackage[english]{babel}
\usepackage{graphicx}
\usepackage{color}
\usepackage{multirow}
\usepackage{booktabs}

\begin{document}

\title{Planar three-body bound states induced by a $p$-wave interatomic resonance}

\author{Maxim A.\ Efremov$^{1,2}$ and Wolfgang P.\ Schleich$^{1,3}$}

\address{$^1$Institut f\"ur Quantenphysik and Center for Integrated Quantum Science and Technology ($\it IQ^{ST}$),
Universit\"at Ulm, Albert-Einstein-Allee 11, D-89081 Ulm, Germany}
\address{$^2$A.M. Prokhorov General Physics Institute, Russian Academy of Sciences, 119991 Moscow, Russia}
\address{$^3$Texas A$\&$M University Institute for Advanced Study (TIAS), Institute for Quantum Science and Engineering (IQSE) and
Department of Physics and Astronomy, Texas A$\&$M University, College Station, Texas 77843-4242, USA}

\eads{\mailto{max.efremov@gmail.com}, \mailto{wolfgang.schleich@uni-ulm.de}}

\begin{abstract}

We consider the bound states of a system consisting of a light particle and two heavy bosonic ones,
which are restricted in their quantum mechanical motion to two space dimensions.
A $p$-wave resonance in the heavy-light short-range potential establishes the interaction between
the two heavy particles.
Due to the large ratio of the atomic masses this planar three-body system is perfectly suited for the Born-Oppenheimer approximation
which predicts a Coulomb energy spectrum with a Gaussian cut-off.

\end{abstract}



\submitto{\NJP}

\maketitle

\section{Introduction}

Borromean rings are an arrangement of three rings linked in such a way that removing one would set the other two free.
In nuclear physics or in cold atoms such three-body bound states are refered to as Efimov states \cite{Efimov}.
However, they occur only when the three particles live in {\it three} space dimensions \cite{Efimov-two dimensions, Jensen},
and the two-particle interaction has a single {\it $s$-wave} resonance \cite{Nishida}.
In the present article we propose a new class of three-body bound states emerging in
a system composed of a light particle and two heavy bosonic ones, when the particles are restricted to {\it two} space
dimensions and the heavy-light short-range interaction potential has a {\it $p$-wave} resonance.
In particular, we show that this situation leads to a Coulomb energy spectrum
with a Gaussian cut-off determined by the large mass ratio of the particles.

\subsection{Interplay between space dimension and symmetry of resonance}

In the case of a three-dimensional $s$-wave resonance, the Efimov states
are supported by a underlying effective potential, which decays with the {\it square} of the distance
\cite{Efimov, Jensen, Nielsen-Fedorov, Braaten-Hammer} between the particles.
However, a three-dimensional $p$-wave resonance leads to an effective potential,
which decays \cite{Efremov-three-dimensions, Zhu Tan} with the {\it cube} of the distance.
Consequently, in the Efimov effect the energy spectrum depends {\it exponentially} \cite{Efimov, Ferlaino-review, He-experiments}
on the quantum number $n$, whereas in the case of the $p$-wave resonance it scales with the {\it sixth power} \cite{Efremov-three-dimensions},
as shown in the first row of Table 1. As a result, in the case of three dimensions, the change of the symmetry of the underlying two-body resonance
results in a {\it finite} number of three-body bound state induced by the $p$-wave resonance, instead of the {\it infinite} number of the Efimov states
originating from the $s$-wave resonance.

We emphasize that the shape of the effective potentials and the associated energy spectra
are intimately connected to the fact that the particles experience a three-dimensional space.
Indeed, in the case of a two-dimensional space there is no Efimov effect \cite{Efimov-two dimensions}. The spectrum of the three-body bound states
is finite and determined by the masses of particles \cite{Belotti} as indicated by the left lower corner of Table 1.

\begin{table}
\caption{\label{tab1} Dependence of the energy $E_n$ for $n\gg 1$ as well as the number $N_b$ of three-body bound states
on the dimensionality of the problem and the symmetry of the underlying two-body resonance.
Here $\alpha$, $\alpha'$, $\beta$, and $n_*$ are constants determined by the masses and quantum statistics of the particles, and
the number of two-body resonant interactions. Moreover, $E_1$ and $E_2$ denote the binding energies of three two-dimensional identical bosons
interacting via a $s$-wave resonant potential expressed in terms of the binding energy $E_b$ of two bosons \cite{Efimov-two dimensions}.
In the case of a $p$-wave resonance
we distinguish the total angular momentum $L=0$ and $L=\pm 1$ discussed in the present article and Ref. \cite{super-Efimov-Moroz}, respectively.}
\begin{indented}
\item[]\begin{tabular}{@{}c|c|c|c}
\hline
\multirow{2}{*}{space} & \multicolumn{2}{r}{symmetry of two-body resonance}\\
\cline{2-4}
dimension & $s$-wave & \multicolumn{2}{c}{$p$-wave} \\ \hline

\multirow{3}{*}{3} & $E_n\propto \exp(-\alpha n)$ & \multicolumn{2}{|c}{$E_n\propto (n_*-n)^6$} \\
& $N_b$ infinite & \multicolumn{2}{|c}{$N_b$ finite} \\
& (Efimov effect) & \multicolumn{2}{|c}{}\\
\hline

\multirow{5}{*}{2} &  & $L=0$ & $L=\pm 1$ \\
\cline{3-4}
& & & \\
& $E_2=-1.27|E_b|$ & $E_n \propto n^{-2}\exp(-\beta n^2)$ & $E_n \propto \exp[-2\exp(\alpha' n)]$ \\
& $E_1=-16.5|E_b|$ &  & \\
& $N_b$ finite & $N_b$ infinite & $N_b$ infinite\\
\hline
\end{tabular}
\end{indented}
\end{table}

In the present article we study a three-body system composed of a light
atom of mass $m$ and two heavy bosonic ones of mass $M$ with a $p$-wave resonance in the heavy-light interaction, provided
the three particles are moving in {\it two} space dimensions.
Due to the large mass ratio the familiar Born-Oppenheimer approximation \cite{LL} allows us to determine the effective potential
by a self-consistent scattering \cite{Efremov-three-dimensions} of the light particle off the two heavy ones in the case of two space dimensions.
The key point of our method \cite{Efremov-three-dimensions} is not to solve the Schr\"odinger equation for the light particle, that is a {\it differential} equation, but
an equivalent system of linear {\it algebraic} equations.

Mixtures with a $p$-wave resonance have already been realized with ${\rm K}$ and ${\rm Rb}$ \cite{P-wave-mixture-KRb},
${\rm Li}$ and ${\rm Rb}$ \cite{P-wave-mixture-LiRb}, as well as with a Rydberg electron and a neutral atom \cite{Rydberg-atoms-p-wave},
corresponding to the mass ratios $m_{\rm K}/M_{\rm Rb}\approx 0.5$, $m_{\rm Li}/M_{\rm Rb}\approx 0.1$, and $m_{\rm e}/M_{\rm Rb}\approx 10^{-5}$, respectively.
Hence, the latter mixtures are promising candidates to verify our predictions.
Moreover, experimentally the reduction of the dimensionality can be achieved by using off-resonant light
to confine ultra-cold gases in one- or two-dimensional periodic lattices \cite{cold-atoms-review}.

In the case of the total angular momentum $L=0$,
we derive an effective potential which is the product of the Efimov potential and the inverse of the logarithm,
with the length scale of the logarithm determined by the parameters of the two-dimensional $p$-wave resonance.
This potential gives rise to an {\it infinite} quasi-Coulomb series
\begin{equation}
 \label{energy spectrum-intoduction}
    E_n=-\frac{E_0}{n^2}\exp\left(-\frac{\pi^2}{2}\frac{\mu}{M}\,n^2\right)
\end{equation}
of three-body bound states for large integer $n$.

As depicted by the left column of the right lower corner of Table 1 and shown on the left of Fig. 1,
the familiar Coulomb series is modified by a Gaussian cut-off governed by the mass ratio $\mu/M$,
where $\mu\equiv 2mM/(m+2M)$ denotes the reduced mass and $E_0$ is the characteristic energy determined
by the short-range physics. In order to observe many states, it is necessary to have the ratio
\begin{equation}
 \label{energy spectrum-ratio}
    \frac{E_{n+1}}{E_n}\cong \exp\left[-\pi^2\frac{\mu}{M}\left(n+\frac{1}{2}\right)\right]
\end{equation}
of the neighboring energies to be of the order of unity, giving rise to the maximum number
\begin{equation}
 \label{n_max}
    n_{\rm max}\cong \frac{1}{\pi^2}\frac{M}{m}
\end{equation}
of observable states as $M\gg m$.

A system consisting of a Rydberg electron and two neutral atoms \cite{Rydberg-atoms-exp, Rydberg-atoms-theory}
with the mass relation $M\cong 10^5 m$, shown on the right panel of Fig. 1, represents a promising candidate to verify our predictions.
In this case, equation (\ref{n_max}) predicts that we can observe
about $10^3$ states of the quasi-Coulomb series given by equation (\ref{energy spectrum-intoduction}).

\begin{figure}
  \centering
   \includegraphics[width=0.45\textwidth]{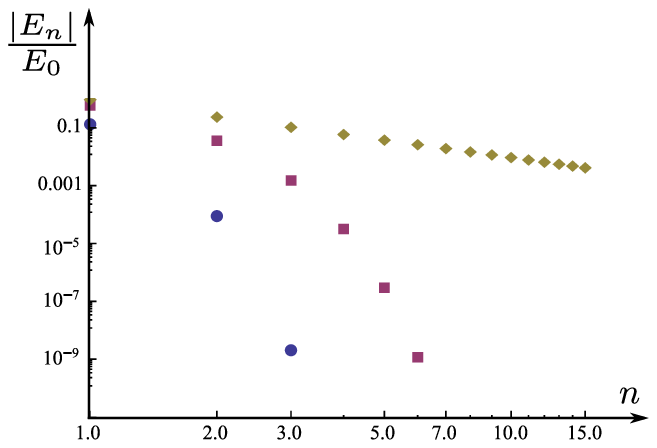}\includegraphics[width=0.5\textwidth]{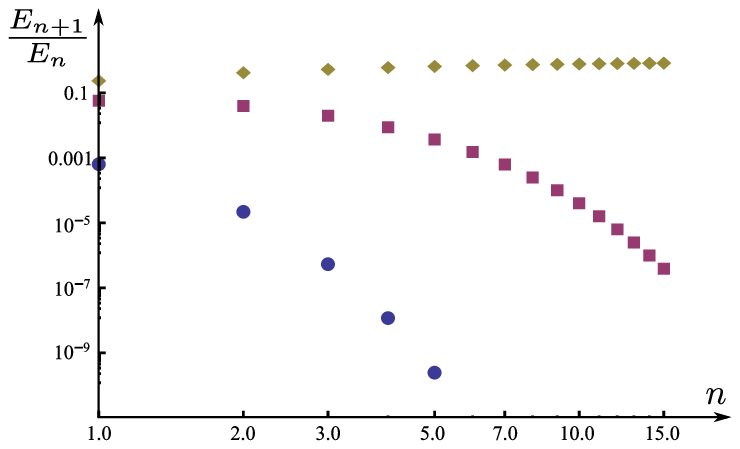}\\
  \caption{Quasi-Coulomb energy spectrum $E_n$  (left) and the resulting ratio $E_{n+1}/E_n$ (right)
  of the neighboring energy eigenvalues given by equation (\ref{energy spectrum-intoduction})
  represented in a double-logarithmic plot for three different mass ratios:
  $m_{\rm K}/M_{\rm Rb}=0.5$ \cite{P-wave-mixture-KRb} (blue dots), $m_{\rm Li}/M_{\rm Rb}=0.1$ \cite{P-wave-mixture-LiRb} (magenta squares),
  $m_{\rm e}/M_{\rm Rb}=10^{-5}$ \cite{Rydberg-atoms-p-wave} (yellow diamonds).}
\end{figure}

In contrast to the case of total angular momentum $L=0$ central to the present discussion,
the situation $L=\pm 1$ analyzed in Ref. \cite{super-Efimov-Moroz} and
summarized in the right column of the right lower corner of Table 1 leads to the ``super Efimov`` effect,
that is the emergence of an {\it infinite} number of the planar three-body bound states
induced by the $p$-wave inter-particle resonance \cite{super-Efimov-Gridnev, super-Efimov-Volosniev}.

\subsection{Outline}

Our article is organized as follows. In section 2 we briefly summarize the essential ingredients of our Born-Oppenheimer
approach towards the three-body problem. We then employ in section 3 a scattering approach to derive a system of linear equations
for the expansion coefficients of the two-dimensional waves. The effective potential for the two heavy particles
follows from the condition that the determinant vanishes providing us with a transcendental equation.
In section 4 we use these relations to rederive the familiar result that in the case of a planar $s$-wave resonance the Efimov effect is absent.
We then dedicate section 5 to a thorough analysis of the effective potentials induced by the planar $p$-wave resonance, and discuss
the resulting bound state spectrum in section 6. Finally section 7 suggests possible ways of observing the predicted bound state series.
We conclude in section 8 by presenting a summary and an outlook.

In order to keep the article self-contained we include an appendix in which we establish a WKB approach towards the time-independent
Schr\"odinger equation corresponding to the induced potentials.
Here we first derive with the help of the Langer transformation \cite{Langer, Loudon}
the exact solution for zero energy, and then obtain approximate analytical expressions for non-zero energies.

\section{Born-Oppenheimer approach}

Since our three-body system consists of a light particle which interacts with two heavy particles,
it can easily be analyzed \cite{Efremov-three-dimensions, Zhu Tan, Fonseca, atomic-molecule}
within the Born-Oppenheimer approximation \cite{LL}.
For this reason, the Schr\"odinger equation
for the full wave function $\Phi({\bf r},{\bf R}) = \Psi({\bf r};{\bf R})\chi({\bf R})$
separates into two two-dimensional equations, and the one for the light particle reads
\begin{equation}
 \label{Schr}
    \left[-\frac{\hbar^2}{2\mu}\Delta_{\bf r}^{(2)}+
    U({\bf r}_{-})+U({\bf r}_{+})\right]\Psi({\bf r};{\bf R})=-\frac{\hbar^2\kappa^2}{2\mu}\Psi({\bf r};{\bf R})
\end{equation}
with ${\bf r}_{\pm}\equiv {\bf r}\pm\frac{1}{2}{\bf R}$. Here ${\bf R}$ denotes the separation between the two heavy particles
and $\Delta_{\bf r}^{(2)}$ is the Laplacian in two dimensions.
For the sake of simplicity we assume the potential $U$ to be cylindrically symmetric
and to have the finite range $r_0$, that is $U(r>R_0)=0$.

The bound-state energies
\begin{equation}
 \label{V-eff}
    {\mathcal V}({\bf R})\equiv -\frac{[\hbar\kappa({\bf R)}]^2}{2\mu}
\end{equation}
of the light particle, corresponding to different expressions for $\kappa$ following from equation (\ref{Schr}),
serve as effective interaction potentials for the relative motion of the heavy particles given by
\begin{equation}
 \label{Schrodinger heavy}
    \left\{\Delta_{\bf R}^{(2)}+\frac{M}{\hbar^2}\left[E-{\mathcal V}({\bf R})\right]\right\}\chi({\bf R})=0,
\end{equation}
where $E$ is the total three-body energy.

For atoms the direct heavy-heavy interaction potential is typically a short-range potential,
and has for large distances a van-der-Waals tail given by $1/R^6$.
We now show that in the case of an exact planar $p$-wave resonance, the effective potential ${\mathcal V}$ is a long-range one,
${\mathcal V}\sim 1/(R^2\ln R)$, and therefore the direct heavy-heavy interaction
has no effect on the behavior of the total potential for large distances.

\section{Interaction potential between heavy particles from scattering theory}

In this section we determine ${\mathcal V}$ by a self-consistent scattering of the light particle off the two heavy ones in two dimensions.
For this purpose, we apply the method suggested in Ref. \cite{Efremov-three-dimensions} and
cast equation (\ref{Schr}) into the integral equation \cite{LL}
\begin{equation}
 \label{Schr integral}
    \Psi({\bf r})=\frac{\mu}{2i\hbar^2}\int d{\bf r}' H_0^{(1)}(i\kappa |{\bf r}-{\bf r}'|)
    \left[U({\bf r}_{-}')+U({\bf r}_{+}')\right]\Psi({\bf r}'),
\end{equation}
where $(-i/4)H_0^{(1)}(i\kappa|{\bf r}|)$ is the Green function of the two-dimensional Schr\"odinger equation (\ref{Schr})
presented in terms of the Bessel function $H_0^{(1)}$ of the third kind, that is the Hankel function \cite{Abramowitz}.

The Schr\"odinger equation (\ref{Schr}) with a two-center potential $U({\bf r}_{-})+U({\bf r}_{+})$ is rather
difficult to work with due to the fact that the solution corresponding to the two potentials is {\it not} a linear superposition
of the two solutions corresponding to a single potential.
However, if we regard the values of the wave function inside the potential wells as given,
the wave function outside is subject to the superposition principle.
Indeed, the heavy-light potential $U$ is a short-range one and
the wave outside of the wells is a solution of the free Schr\"odinger equation in the form of a purely outgoing wave.
Since the total heavy-light potential $U({\bf r}_{-})+U({\bf r}_{+})$
is nonzero only inside two spheres of radius $r_0$ centered at ${\bf r}=\pm\frac{1}{2}{\bf R}$,
we represent equation (\ref{Schr integral}) as the superposition
\begin{equation}
 \label{solution-superposition}
    \Psi({\bf r})=\Psi^{(-)}({\bf r})+ \Psi^{(+)}({\bf r})
\end{equation}
of the two waves
\begin{equation}
 \label{Psi-sigma definition}
    \Psi^{(\pm)}({\bf r})\equiv\int\limits_{|{\bf r}'\pm\frac{{\bf R}}{2}|\leq r_0}d{\bf r}'
    \sigma ^{(\pm)}({\bf r}')H_0^{(1)}(i\kappa |{\bf r}-{\bf r'}|)
\end{equation}
with
\begin{equation}
 \label{sigma}
    \sigma ^{(\pm)}({\bf r})\equiv \frac{\mu}{2i\hbar^2}\,U\left({\bf r}\pm \frac{1}{2}{\bf R}\right)\Psi({\bf r}).
\end{equation}

The addition theorem \cite{Abramowitz}
\begin{equation}
 \label{adition theorem}
   H_0^{(1)}(i\kappa|{\bf r}-{\bf r'}|)=
   \sum_{m=-\infty}^{\infty}J_m(i\kappa r')H_m^{(1)}(i\kappa r)e^{im(\varphi_{\bf r}-\varphi_{\bf r'})}
\end{equation}
for the Bessel functions $J_m$ and $H_m^{(1)}$ with $m=0,\pm 1,\pm 2$,.. and $r>r'$,
transforms equation (\ref{Psi-sigma definition}) into
\begin{equation}
 \label{Psi sums}
   \Psi^{(\pm)}({\bf r})=\frac{1}{\sqrt{2\pi}}\sum_{m=-\infty}^{\infty}C_{m}^{(\pm)}H_m^{(1)}(i\kappa r_{\pm})e^{im\varphi_{\bf r_{\pm}}},
\end{equation}
with $\varphi_{\bf r}$ being the polar angle  of the two-dimensional vector ${\bf r}\equiv(r\cos\varphi_{\bf r},r\sin\varphi_{\bf r})$.

We regard the coefficients $C_m^{(\pm)}$ determined by the integral in equation (\ref{Psi-sigma definition}) as independent variables
and apply scattering theory to obtain from equations (\ref{solution-superposition}) and (\ref{Psi sums}) explicit equations for $C_m^{(\pm)}$
coupled by the $S$-matrix elements of the potential $U$.
For this purpose, we consider a vicinity of the first potential well, that is ${\bf r}=-\frac{1}{2}{\bf R}+{\bf x}$
with $|{\bf x}|\approx r_0$, where the total solution
$$
\Psi\left(-\frac{1}{2}{\bf R}+{\bf x}\right)=\frac{1}{\sqrt{2\pi}}\sum_{m=-\infty}^{\infty}R_m(|{\bf x}|)e^{im\varphi_{\bf x}}
$$
given by equation (\ref{solution-superposition}) can be expanded into the polar harmonics $e^{im\varphi_{\bf x}}$.
Here the radial wave function
\begin{equation}
 \label{radial function}
   R_m(|{\bf x}|)=C_m^{(+)}H_m^{(1)}(i\kappa |{\bf x}|)+J_m(i\kappa |{\bf x}|)
  \sum_{m'=-\infty}^{\infty}H_{m-m'}^{(1)}(i\kappa R)e^{i(m'-m)\varphi_{\bf R}}C_{m'}^{(-)}
\end{equation}
is determined by the sum of the two contributions resulting from $\Psi^{(\pm)}(-\frac{1}{2}{\bf R}+{\bf x})$
defined by equation (\ref{Psi sums}), and the sum over $m'$ originates from the re-expansion of
$H_m^{(1)}(i\kappa|{\bf x}-{\bf R}|)e^{im\varphi_{{\bf x}-{\bf R}}}$ into the polar harmonics $e^{im\varphi_{\bf x}}$.

In order to derive an equation for $C_m^{(\pm)}$ we cast the
radial wave $R_m$ given by equation (\ref{radial function}) into the superposition
$$
R_m(|{\bf x}|)=a_m(\kappa)H_m^{(1)}(i\kappa |{\bf x}|)+b_m(\kappa)H_m^{(2)}(i\kappa |{\bf x}|)
$$
of {\it outgoing} and {\it incoming} radial waves $H_m^{(1)}$ and $H_m^{(2)}$ with amplitudes
\begin{equation}
 \label{a coefficient}
  a_m=C_m^{(+)}+\frac{1}{2}\sum_{m'=-\infty}^{\infty}H_{m-m'}^{(1)}(i\kappa R)e^{i(m'-m)\varphi_{\bf R}}C_{m'}^{(-)}
\end{equation}
and
\begin{equation}
 \label{b coefficient}
  b_m=\frac{1}{2}\sum_{m'=-\infty}^{\infty}H_{m-m'}^{(1)}(i\kappa R)e^{i(m'-m)\varphi_{\bf R}}C_{m'}^{(-)}.
\end{equation}
Here $J_m$ is expressed \cite{Abramowitz} in terms of the Bessel functions of the third kind
$H_m^{(1)}$ and $H_m^{(2)}$ as $J_m(z)=[H_m^{(1)}(z)+H_m^{(2)}(z)]/2$.

Since the amplitudes $a_m$ and $b_m$ of the {\it outgoing} and {\it incoming} waves
are coupled \cite{LL, Newton} by the on-shell $T$-matrix elements
\begin{equation}
 \label{S-matrix-phase}
  T_m(i\kappa)=-\frac{1}{\pi}\frac{1}{\cot[\delta_m(i\kappa)]-i}
\end{equation}
of the scattering potential $U$, with $\delta_m$ being the corresponding scattering phase, that is
$$
\frac{a_m(\kappa)}{b_m(\kappa)}=1-2i\pi\,T_m(i\kappa),
$$
we arrive at
\begin{equation}
 \label{C equation 1}
    C_m^{(+)}+i\pi\,T_m(i\kappa)\sum_{m'=-\infty}^{\infty}H_{m-m'}^{(1)}(i\kappa R)C_{m'}^{(-)}=0.
\end{equation}
Here we have used the fact that the vector ${\bf R}$ is directed along the $x$-axis, and hence $\varphi_{\bf R}=0$.

Similarly, we obtain from the second potential well centered at ${\bf r}=\frac{1}{2}{\bf R}$, the relation
\begin{equation}
 \label{C equation 2}
    C_m^{(-)}+i\pi\,T_m(i\kappa)\sum_{m'=-\infty}^{\infty}H_{m'-m}^{(1)}(i\kappa R)C_{m'}^{(+)}=0.
\end{equation}

The values of $\kappa=\kappa(R)$, and consequently the effective potential ${\mathcal V}$ given by equation (\ref{V-eff}),
follow \cite{Efremov-three-dimensions} from the condition that
the determinant of the system of linear algebraic equations (\ref{C equation 1})-(\ref{C equation 2})
for the coefficients $C_m^{(\pm)}$ vanishes. This constraint provides us with a transcendental equation for $\kappa$.

We emphasize that a similar set of equations emerges \cite{Efremov-three-dimensions} in the three-dimensional case.
However, the main difference in two dimensions is the appearance of the Hankel functions $H_m^{(1)}$ with an integer
rather than a half-integer index.
This substitution is a consequence of the reduced dimensionality and can be traced back to the Green function \cite{LL}
of a free particle in two dimensions being proportional to $H_0^{(1)}$.

\section{Planar $s$-wave resonance: No Efimov effect}

In this section we apply our method to the case when the heavy-light interaction potential $U$ is of zero-range with a single $s$-wave resonance.
This technique allows us to rederive immediately the results of Ref. \cite{Belotti}
and serves as a guide for our next move to the uncharted territory of planar $p$-wave resonances analyzed in the following section.

The $s$-wave resonance corresponds to $m=0$ and the system of equations (\ref{C equation 1}) and (\ref{C equation 2}) reduces to
\begin{eqnarray}
    C_0^{(+)}+i\pi\,T_0(i\kappa)H_{0}^{(1)}(i\kappa R)C_{0}^{(-)}=0 \nonumber \\
    C_0^{(-)}+i\pi\,T_0(i\kappa)H_{0}^{(1)}(i\kappa R)C_{0}^{(+)}=0. \nonumber
\end{eqnarray}

The condition of a vanishing determinant leads us to the relation
\begin{equation}
 \label{zero-range k-equation}
 i\pi\,T_0(i\kappa)H_{0}^{(1)}(i\kappa R)=\pm 1.
\end{equation}

In the low-energy limit, that is for $\kappa r_0\ll 1$, the relation \cite{two-dimensional scattering theory}
\begin{equation}
 \label{zero-range-appendix}
    \cot[\delta_0(i\kappa)]\cong\frac{2}{\pi}\left[\gamma+\ln\left(\frac{i\kappa a_0}{2}\right)\right]
\end{equation}
involves the two-dimensional scattering length $a_0$ linked to the energy
\begin{equation}
 \label{zero-range-bound-energy}
    \varepsilon_0\equiv-2e^{-2\gamma}\frac{\hbar^2}{\mu a_0^2}
\end{equation}
of the $s$-wave weakly-bound state in the heavy-light potential $U$ for $a_0\gg r_0$.
Here $\gamma$ is the Euler constant.

When we substitute equations (\ref{S-matrix-phase}) and (\ref{zero-range-appendix}) into the condition equation (\ref{zero-range k-equation}),
recall the identity $K_m(z)=(\pi/2)i^{m+1}H_m^{(1)}(iz)$ for the modified Bessel function of the second kind \cite{Abramowitz}, and
introduce the dimensionless variable $\xi\equiv e^{\gamma}(\kappa a_0/2)$, we arrive at the transcendental equation
\begin{equation}
 \label{energy-eq-Efimov}
    K_0\left[\left(\frac{2}{e^{\gamma}}\frac{R}{a_0}\right)\xi\right]=\pm\ln\xi.
\end{equation}

The two solutions $\xi_{\pm}$ of equation (\ref{energy-eq-Efimov}) determine the two effective potentials
\begin{equation}
 \label{energy-m0-definition}
    \mathcal{V}_{\pm}(R)\equiv \varepsilon_0\xi_{\pm}^2(R),
\end{equation}
shown in Fig. 2, which for large and small distances $R$ read
\begin{equation}
 \label{energy-asymptotic-m0-symmetric}
    {\mathcal V}_{+}(R)=-|\varepsilon_0|
    \left\{
      \begin{array}{ll} \frac{a_0}{R}\;\;\;\;\;\;\;\;\;\;\;\;\;\;\;\;\;\;\;\;\;\;\;\;\;\;\;\;\;\;\;\;\;\;\;\;\;\, {\rm for}\; R\ll a_0\\
    \\ 1+\sqrt{\pi e^{\gamma}\frac{a_0}{R}}\exp\left(-\frac{2}{e^{\gamma}}\frac{R}{a_0}\right)\;\; {\rm for}\; R\gg a_0
      \end{array}\right.
\end{equation}
and
\begin{equation}
 \label{energy-asymptotic-m0-asymmetric}
    {\mathcal V}_{-}(R)=-|\varepsilon_0|
    \left\{
      \begin{array}{ll} e^{2\gamma}\ln\left(\frac{R}{a_0}\right)\frac{a_0^2}{R^2}\;\;\;\;\;\;\;\;\;\;\;\;\;\;\;\;\;\;\;\;\;{\rm for}\;R\gtrsim a_0\\
    \\ 1-\sqrt{\pi e^{\gamma}\frac{a_0}{R}}\exp\left(-\frac{2}{e^{\gamma}}\frac{R}{a_0}\right)\;\; {\rm for}\; R\gg a_0.
      \end{array}\right.
\end{equation}

As depicted in Fig. 2, the potential ${\mathcal V}_{-}$ is repulsive for $R\geq a_0$.
Here we do not show ${\mathcal V}_{-}$ for $R<a_0$ since in this regime the solution $\xi_{-}$ of equation (\ref{energy-eq-Efimov}) is complex-valued.

In contrast to ${\mathcal V}_{-}$ the potential ${\mathcal V}_{+}$ is attractive and has the range of the order of $a_0$.
Thus, the potential ${\mathcal V}_{+}$ can support only a finite number of three-body bound states,
resulting in the absence \cite{Efimov-two dimensions, Belotti} of the Efimov effect in two dimensions.

\begin{figure}
  \centering
   \includegraphics[width=0.75\textwidth]{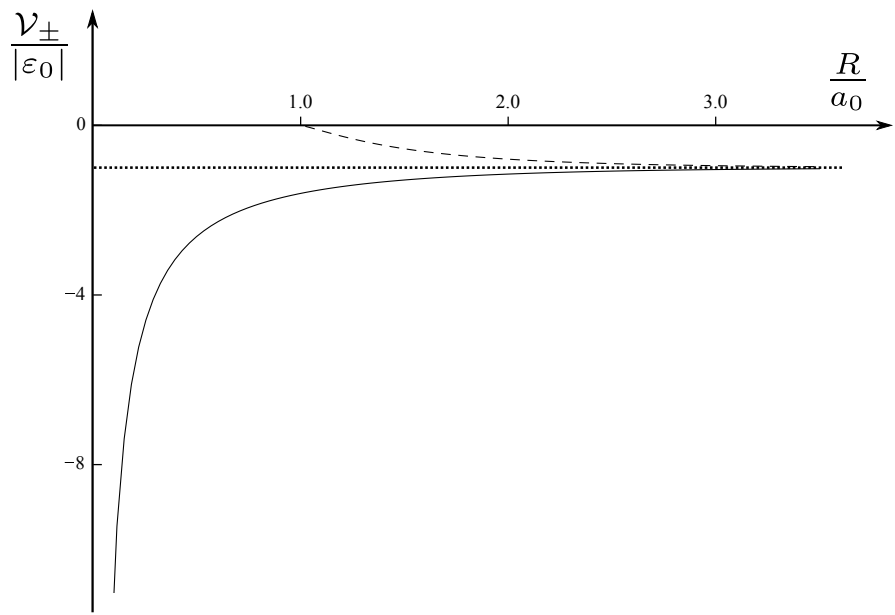}\\
  \caption{Effective potentials ${\mathcal V}_{+}={\mathcal V}_{+}(R)$ (solid curve) and ${\mathcal V}_{-}={\mathcal V}_{-}(R)$ (dashed curve)
  induced by a $s$-wave resonant two-body interaction in two space dimensions and given by equations (\ref{energy-eq-Efimov}) and (\ref{energy-m0-definition})
  as functions of the separation $R$ between the two heavy particles.
  Since ${\mathcal V}_{-}$ is repulsive, three-bound states can only result from the attractive potential ${\mathcal V}_{+}$.
  However, due to the exponential decay of ${\mathcal V}_{+}(R)$ towards $-|\varepsilon_0|$ for $R\rightarrow \infty$,
  as shown by equation (\ref{energy-asymptotic-m0-symmetric}), only a {\it finite} number of three-body bound states emerges.
  As a consequence, the Efimov effect does not exist in two space dimensions.}
\end{figure}

\section{Planar $p$-wave resonance: Emergence of new three-body bound states}

In this section we address the case of a $p$-wave resonance in the heavy-light potential $U$ and
first derive the transcendental equations for $\kappa$ starting from the system of equations (\ref{C equation 1}) and (\ref{C equation 2}).
We then solve these equations for $\kappa$
to obtain the effective potentials ${\mathcal V}$ given by equation (\ref{V-eff}).

\subsection{Derivation of transcendental equations for $\kappa$}

Similar to the expression, equation (\ref{zero-range-appendix}), for the $s$-wave scattering phase,
in the low-energy limit, that is for $\kappa r_0\ll 1$,
we can again parametrize the corresponding formula for the $p$-wave scattering phase
within the two-dimensional effective-range expansion \cite{effective range} by
the $p$-wave effective scattering length $a_1$ and effective range $r_1$, leading us to
\begin{equation}
 \label{p-wave-low-energy}
  \cot[\delta_1(i\kappa)]\cong \frac{2}{\pi}\left[\frac{1}{a_1 \kappa^2}+\ln\left(i\kappa r_1\right)\right].
\end{equation}

Here $a_1$ and $r_1$ are non-negative parameters and for $a_1\gg r_1^2$ they are determined by the energy
\begin{equation}
 \label{energy-p-wave}
  \varepsilon_1\cong -\frac{\hbar^2}{\mu a_1\ln[a_1/(2 r_1^2)]}
\end{equation}
of the $p$-wave bound state.

The latter is defined by the positive-valued pole $\kappa_1$ of the matrix element $T_1$
given by equation (\ref{S-matrix-phase}) with equation (\ref{p-wave-low-energy}), that is
\begin{equation}
 \label{energy-p-wave-defnition}
  \varepsilon_1\equiv -\frac{\hbar^2\kappa_1^2}{2\mu}.
\end{equation}

Moreover, in the case of a $p$-wave resonance, that is $a_1^{-1}=0$,
the effective range $r_1$ has an upper bound \cite{effective range} determined by the potential range $r_0$,
that is $r_1\leq \frac{1}{2} e^{\gamma}r_0$, for any two-dimensional short-range potential.

In the case of  $a_1^{-1}=0$ the matrix element $T_1$ given by equations (\ref{S-matrix-phase}) and (\ref{p-wave-low-energy}) for the resonant channel
is of the same order as $T_{0}$ determined by equations (\ref{S-matrix-phase}) and (\ref{zero-range-appendix}) for the non-resonant channel.
Moreover, due to the scaling $T_{m>1}(i\kappa)\sim (\kappa r_0)^{2m}$ in the low-energy limit,
we need to take into account in equations (\ref{C equation 1}) and (\ref{C equation 2}) only the $s$- and $p$-waves, giving rise
to the system
\begin{equation}
 \label{p-wave k-equations}
 \begin{array}{r@{}l}
  C_0^{(+)}+\alpha_1 C_{-1}^{(-)}+\alpha_0 C_0^{(-)}-\alpha_1 C_1^{(-)}=0 \\
  C_0^{(-)}-\alpha_1 C_{-1}^{(+)}+\alpha_0 C_0^{(+)}+\alpha_1 C_{1}^{(+)}=0 \\
  C_1^{(+)}+\beta_2 C_{-1}^{(-)}+\beta_1 C_0^{(-)}+\beta_0 C_1^{(-)}=0 \\
  C_{-1}^{(+)}+\beta_0 C_{-1}^{(-)}-\beta_1 C_0^{(-)}+\beta_2 C_1^{(-)}=0 \\
  C_1^{(-)}+\beta_2 C_{-1}^{(+)}-\beta_1 C_0^{(+)}+\beta_0 C_1^{(+)}=0 \\
  C_{-1}^{(-)}+\beta_0 C_{-1}^{(+)}+\beta_1 C_0^{(+)}+\beta_2 C_1^{(+)}=0
 \end{array}
\end{equation}
of six algebraic equations for $C_{0}^{(\pm)}$, $C_{1}^{(\pm)}$, and $C_{-1}^{(\pm)}$.
Here we have defined the abbreviations
\begin{equation}
 \label{alpha coefficients}
  \alpha_{m}\equiv i\pi T_0(i\kappa)H_{m}^{(1)}(i\kappa R)
\end{equation}
for $m=0,1$ and
\begin{equation}
 \label{beta coefficients}
  \beta_{m'}\equiv i\pi T_1(i\kappa)H_{m'}^{(1)}(i\kappa R)
\end{equation}
for $m'=0,1,2$.

In order to find a non-trivial solution of equation (\ref{p-wave k-equations}),
we cast equation (\ref{p-wave k-equations}) into the form of the linear equation
\begin{equation}
 \label{p-wave-matrix form}
  {\mathcal M}{\bf C}=0
\end{equation}
for the vector
\begin{equation}
 \label{c-vector}
  {\bf C}\equiv
    \begin{pmatrix}
   C_0^{(+)}+C_0^{(-)}\\C_1^{(+)}-C_{-1}^{(+)}-C_1^{(-)}+C_{-1}^{(-)}\\C_0^{(+)}-C_0^{(-)}\\
   C_1^{(+)}-C_{-1}^{(+)}+C_1^{(-)}-C_{-1}^{(-)}\\C_1^{(+)}+C_{-1}^{(+)}\\C_1^{(-)}+C_{-1}^{(-)}
  \end{pmatrix},
\end{equation}
where the matrix
\begin{equation}
 \label{matrix M}
  {\mathcal M}\equiv
  \begin{pmatrix}
   {\mathcal M}_+ & 0 & 0 \\
   0 & {\mathcal M}_- & 0 \\
   0 & 0 & {\mathcal M}_0
  \end{pmatrix}
\end{equation}
has the block-diagonal form, with
\begin{equation}
 \label{matrix M1}
  {\mathcal M}_{\pm}\equiv
  \begin{pmatrix}
   1\pm\alpha_0 & \pm\alpha_1 \\
   \pm 2\beta_1 & 1\pm(\beta_2-\beta_0)
  \end{pmatrix}
\end{equation}
and
\begin{equation}
 \label{matrx M3}
  {\mathcal M}_0\equiv
  \begin{pmatrix}
   1 & \beta_0+\beta_2 \\
   \beta_0+\beta_2 & 1
  \end{pmatrix}.
\end{equation}

As a result, equation (\ref{p-wave-matrix form}) has a non-trivial solution
only if the determinant of ${\mathcal M}$ vanishes, giving rise to the two conditions
\begin{equation}
 \label{p-wave condition 1}
  \beta_0+\beta_2=\mp 1,
\end{equation}
and
\begin{equation}
 \label{p-wave condition 2}
  (\alpha_0\pm 1)(\beta_2-\beta_0\pm 1)-2\alpha_1\beta_1=0.
\end{equation}

When we substitute equations (\ref{S-matrix-phase}) and (\ref{p-wave-low-energy}) into the condition equation (\ref{p-wave condition 1}) and
recall the definition equation (\ref{beta coefficients}) with the identity \cite{Abramowitz} $K_m(z)=(\pi/2)i^{m+1}H_m^{(1)}(iz)$ ,
we arrive at the transcendental equation
\begin{equation}
 \label{p-wave condition 1-modified}
  K_0\left(\xi\,\frac{R}{r_1}\right)-K_2\left(\xi\,\frac{R}{r_1}\right)=\pm\left(\frac{r_1^2}{a_1}\frac{1}{\xi^2}+\ln\xi\right)
\end{equation}
for the dimensionless variable $\xi\equiv\kappa r_1$.

The two solutions $\xi_{\rm I}^{(\pm)}$ of equation (\ref{p-wave condition 1-modified}) correspond to the wave function $\Psi=\Psi({\bf r})$
of the light particle, equations (\ref{solution-superposition}) and (\ref{Psi sums}), having the form
of the symmetric, $\Psi_{{\rm I}}^{(+)}$, and asymmetric, $\Psi_{{\rm I}}^{(-)}$, superpositions of the pure $p$-waves,
that is
\begin{equation}
 \label{Psi-I}
  \Psi_{{\rm I}}^{(\pm)}({\bf r})\propto \left[\sin(\varphi_{{\bf r}_+})K_1(\kappa
r_+)\pm \sin(\varphi_{{\bf r}_-})K_1(\kappa r_-)\right].
\end{equation}
Here the subscript ${\rm I}$ refers to the first branch.

Similarly, we can rewrite the second condition, equation (\ref{p-wave condition 2}), in the form of the transcendental equation
$$
\left[K_2\left(\xi\frac{R}{r_1}\right)+K_0\left(\xi\frac{R}{r_1}\right)\pm\left(\frac{r_1^2}{a_1}\frac{1}{\xi^2}+\ln\xi\right)\right]\times
$$
\begin{equation}
 \label{p-wave condition 2-modified}
  \left[K_0\left(\xi\frac{R}{r_1}\right)\mp\ln\left(\xi\frac{e^{\gamma}}{2}\frac{a_0}{r_1}\right)\right]=2K_1^2\left(\xi\frac{R}{r_1}\right).
\end{equation}

The two solutions $\xi_{\rm II}^{(\pm)}$ of equation (\ref{p-wave condition 2-modified})
correspond to the wave-function $\Psi=\Psi({\bf r})$ of the light
particle, equations (\ref{solution-superposition}) and (\ref{Psi sums}), having the form of the symmetric,
$\Psi_{{\rm II}}^{(+)}$, and asymmetric, $\Psi_{{\rm II}}^{(-)}$, superpositions
of the $s$- and $p$-waves, that is
\begin{equation}
 \label{Psi-II}
  \Psi_{{\rm II}}^{(\pm)}({\bf r})\propto
K_0(\kappa r_{+})\pm K_0(\kappa r_{-})-A_{\pm}[\cos(\varphi_{{\bf r}_+})K_1(\kappa
r_+)\mp\cos(\varphi_{{\bf r}_{-}})K_1(\kappa r_-)],
\end{equation}
with $A_{\pm}\equiv[2T_0K_0(\kappa R)\pm 1][2T_0K_1(\kappa R)]^{-1}$. Here the subscript ${\rm II}$ refers to the second branch.

\subsection{Solutions of two transcendental equations for $\kappa$}

Next we focus on solving the two transcendental equations (\ref{p-wave condition 1-modified}) and (\ref{p-wave condition 2-modified})
for $\kappa$ in the case of an exact $p$-wave resonance, that is for $a_1^{-1}=0$,
and obtaining the corresponding effective potentials ${\mathcal V}$, equation (\ref{V-eff}).
We first address equation (\ref{p-wave condition 1-modified}) and
then turn to the slightly more complicated case of equation (\ref{p-wave condition 2-modified}).

\subsubsection{Solution of equation (\ref{p-wave condition 1-modified}): superposition of $p$-waves}

When we introduce the variable $z\equiv \xi (Rr_1^{-1})=\kappa R$ and
use the asymptotic behavior \cite{Abramowitz} of $K_m(z)$ as $z\rightarrow 0$ with $m=0$ and $m=2$,
we cast equation (\ref{p-wave condition 1-modified}) for the case of $a_1^{-1}=0$ into the form
\begin{equation}
 \label{condition 1-expansion}
  K_0(z)-K_2(z)\cong -\frac{2}{z^2}-\ln\frac{z}{2}-\gamma+\frac{1}{2}+\mathcal{O}\left(z^2\right)=\pm\left(\ln z-\ln\frac{R}{r_1}\right),
\end{equation}
which has a solution only for the positive sign on the right-hand side as $R\gg r_1$ and $z\ll 1$.

Moreover, we introduce the variable $\rho\equiv \exp(\frac{1}{2}-\gamma)(Rr_1^{-1})$, rewrite equation (\ref{condition 1-expansion}) as
$$
\frac{2}{z^2}=\ln\rho+\ln\frac{2}{z^2},
$$
and solve this equation perturbativally with respect to the term $\ln(z^2/2)$ as $\rho\gg 1$.

As a result, we arrive at the solution
$$
z^2\cong \frac{2}{\ln\rho+\ln\ln\rho}
$$
or
$$
\left(\xi_I^{(0)}\right)^2\cong \frac{2r_1^2}{R^2}\frac{1}{\ln\frac{R}{r_1}-\gamma+\frac{1}{2}+\ln\left(\ln\frac{R}{r_1}-\gamma+\frac{1}{2}\right)}.
$$

Thus, the effective potential
\begin{equation}
 \label{energy-asymptotic-first-definition}
    {\mathcal V}_{{\rm I}}^{(0)}\equiv -\frac{\hbar^2}{2\mu r_1^2}\left(\xi_{{\rm I}}^{(0)}\right)^2
\end{equation}
given by equation (\ref{V-eff}) reads
\begin{equation}
 \label{energy-asymptotic-first}
    {\mathcal V}_{{\rm I}}^{(0)}\cong
    -\frac{\hbar^2}{\mu R^2}\frac{1}{\ln\frac{R}{r_1}-\gamma+\frac{1}{2}+\ln\left(\ln\frac{R}{r_1}-\gamma+\frac{1}{2}\right)}
\end{equation}
for $R\gg r_1$.

However, near the resonance, $a_1\gg r_1^2$, that is in the
case of the weakly-bound $p$-wave state in $U$, the effective potentials
\begin{equation}
 \label{energy-asymptotic-pm-first}
    {\mathcal V}_{{\rm I}}^{(\pm)}\equiv -\frac{\hbar^2}{2\mu r_1^2}\left(\xi_{{\rm I}}^{(\pm)}\right)^2
\end{equation}
are determined by the two solutions $\xi_{{\rm I}}^{(\pm)}$ of equation (\ref{p-wave condition 1-modified}) and presented in Fig. 3.
The ranges of ${\mathcal V}_{{\rm I}}^{(+)}$ and ${\mathcal V}_{{\rm I}}^{(-)}$ are identical and equal to
\begin{equation}
 \label{range R_1}
    R_1\equiv\frac{\hbar}{\sqrt{2\mu|\varepsilon_1|}}\cong \left(\frac{a_1}{2}\ln\frac{a_1}{2r_1^2}\right)^{\frac{1}{2}}.
\end{equation}

For large distances, $R\gtrsim R_1$, they approach exponentially the bound state energy $\varepsilon_1$ of the light particle
defined by equation (\ref{energy-p-wave}), as depicted in Fig. 3. For short distances, $R\lesssim R_1$,
the potential ${\mathcal V}_{\rm I}^{(+)}$ decreases monotonically and approaches ${\mathcal V}_{\rm I}^{(0)}$, equation
(\ref{energy-asymptotic-first}), whereas ${\mathcal V}_{\rm I}^{(-)}$ increases monotonically and vanishes at $R=\sqrt{2a_1}$.

\begin{figure}
  \centering
   \includegraphics[width=0.9\textwidth]{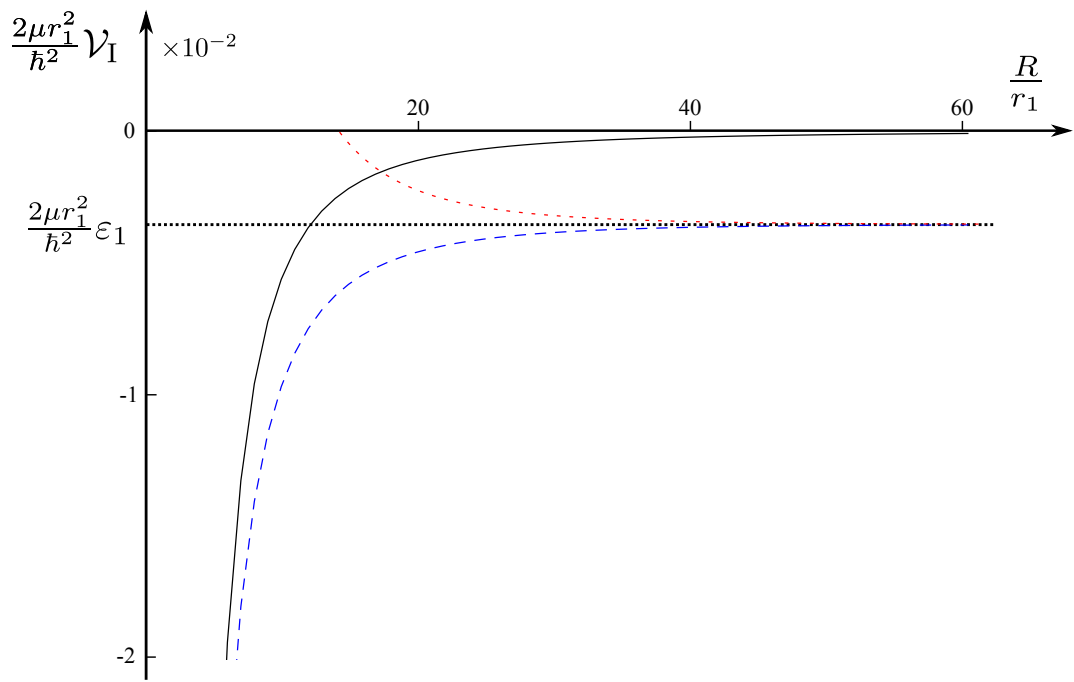}\\
  \caption{Effective potentials ${\mathcal V}_{\rm I}^{(0)}={\mathcal V}_{\rm I}^{(0)}(R)$ (solid black curve),
  ${\mathcal V}_{\rm I}^{(+)}={\mathcal V}_{\rm I}^{(+)}(R)$ (dashed blue curve),
  and ${\mathcal V}_{\rm I}^{(-)}={\mathcal V}_{\rm I}^{(-)}(R)$ (dotted red curve)
  induced by a $p$-wave resonant two-body interaction in two space dimensions and corresponding to the first branch,
  equations (\ref{energy-asymptotic-first-definition}) and (\ref{energy-asymptotic-pm-first}),
  as functions of the separation $R$ between the two heavy particles. Here $\varepsilon_1$ is the energy of the $p$-wave bound state
  given by equations (\ref{energy-p-wave}) and (\ref{energy-p-wave-defnition})
  for the $p$-wave scattering length $a_1=100\,r_1^2$, with $r_1$ being the $p$-wave effective range.
  In complete accordance with the case of a $s$-wave resonant interaction, discussed in Fig. 2,
  we again find two potentials ${\mathcal V}_{\rm I}^{(\pm)}$, one attractive and one repulsive.
  However, in contrast to a $s$-wave resonance we can now have $\varepsilon_1=0$ corresponding to an exact $p$-wave resonance
  giving rise to ${\mathcal V}_{\rm I}^{(0)}$. Only this potential decreases appropriately for $R\rightarrow \infty$ as to support
  {\it infinitely} many three-body bound states.}
\end{figure}

\subsubsection{Solution of equation (\ref{p-wave condition 2-modified}): superposition of $s$- and $p$-waves}

Finally we turn to the second branch of solutions of the truncated system of equations (\ref{p-wave k-equations})
determined by the transcendental equation (\ref{p-wave condition 2-modified}).

In the case of $a_1^{-1}=0$, we introduce again the variable $z\equiv \xi (Rr_1^{-1})=\kappa R$ and
use the asymptotic behavior \cite{Abramowitz} of $K_m(z)$ as $z\rightarrow 0$ with $m=0,1,2$,
which leads us to the form
\begin{equation}
 \label{condition 2-expansion}
  \left[\ln\frac{R}{r_1}\mp a_{\pm}\right]\left[\ln\frac{R}{r_1}\pm b_{\pm}\right]=-\frac{2}{z^2}-\ln\frac{z^2}{4}-2\gamma+1+\mathcal{O}(z^2)
\end{equation}
of equation (\ref{p-wave condition 2-modified}) with the asymptotic expressions for the functions
$$
a_{+}\cong \frac{2}{z^2}-\gamma-\frac{1}{2}+\ln2+\mathcal{O}(z^2)
$$
$$
a_{-}\cong \frac{2}{z^2}-\ln\frac{z^2}{2}-\gamma-\frac{1}{2}+\mathcal{O}(z^2)
$$
$$
b_{+}\cong -2\gamma-\ln\frac{a_0}{r_1}-\ln\frac{z^2}{4}-\frac{z^2}{4}\left(\ln\frac{z}{2}+\gamma-1\right)+\mathcal{O}(z^2)
$$
$$
b_{-}\cong \ln\frac{a_0}{r_1}-\frac{z^2}{4}\left(\ln\frac{z}{2}+\gamma-1\right)+\mathcal{O}(z^2)
$$
as $z$ tends to zero.

A solution of equation (\ref{condition 2-expansion}) occurs only with non-vanishing values of $a_{+}$ and $b_{+}$.
In order to find this solution we rewrite this equation as
$$
\ln\frac{R}{r_1}=\frac{2}{z^2}-\gamma-\frac{1}{2}+\ln2-\frac{2+\mathcal{O}(z^2)}{z^2\left(\ln\frac{R}{r_1}\right)+\mathcal{O}(z^2)}
$$
and solve it perturbativally with respect to the last term on the right-hand side, resulting in
$$
\ln\frac{R}{r_1}\cong\frac{2}{z^2}-\gamma-\frac{3}{2}+\ln2+\mathcal{O}(z^2)
$$
or
$$
\left(\xi_{II}^{(0)}\right)^2\cong\frac{2r_1^2}{R^2}\frac{1}{\ln\frac{R}{2r_1}+\gamma+\frac{3}{2}}.
$$

Thus, the effective potential
\begin{equation}
 \label{energy-asymptotic-second-definition}
    {\mathcal V}_{{\rm II}}^{(0)}\equiv -\frac{\hbar^2}{2\mu r_1^2}\left(\xi_{{\rm II}}^{(0)}\right)^2
\end{equation}
given by equation (\ref{V-eff}) reads
\begin{equation}
 \label{energy-asymptotic-second}
    {\mathcal V}_{{\rm II}}^{(0)}\cong
    -\frac{\hbar^2}{\mu R^2}\frac{1}{\ln\frac{R}{2r_1}+\gamma+\frac{3}{2}}
\end{equation}
for $R\gg r_1$.

As a result, on the $p$-wave resonance and for $R\rightarrow \infty$, both potentials ${\mathcal V}_{\rm I}^{(0)}$ and ${\mathcal V}_{\rm II}^{(0)}$
given by equations (\ref{energy-asymptotic-first}) and (\ref{energy-asymptotic-second}) approach the same behavior
\begin{equation}
 \label{V asymptotic}
  {\mathcal V_{\rm I}^{(0)}}(R\rightarrow \infty)\cong {\mathcal V_{\rm II}^{(0)}}(R\rightarrow \infty)\cong
  {\mathcal V}(R)\equiv -\frac{\hbar^2}{\mu R^2}\frac{1}{\ln\frac{R}{r_1}}
\end{equation}
determined by the $p$-wave effective range $r_1$, where $r_1$ is of the order of the two-body range $r_0$.

\begin{figure}
  \centering
   \includegraphics[width=0.9\textwidth]{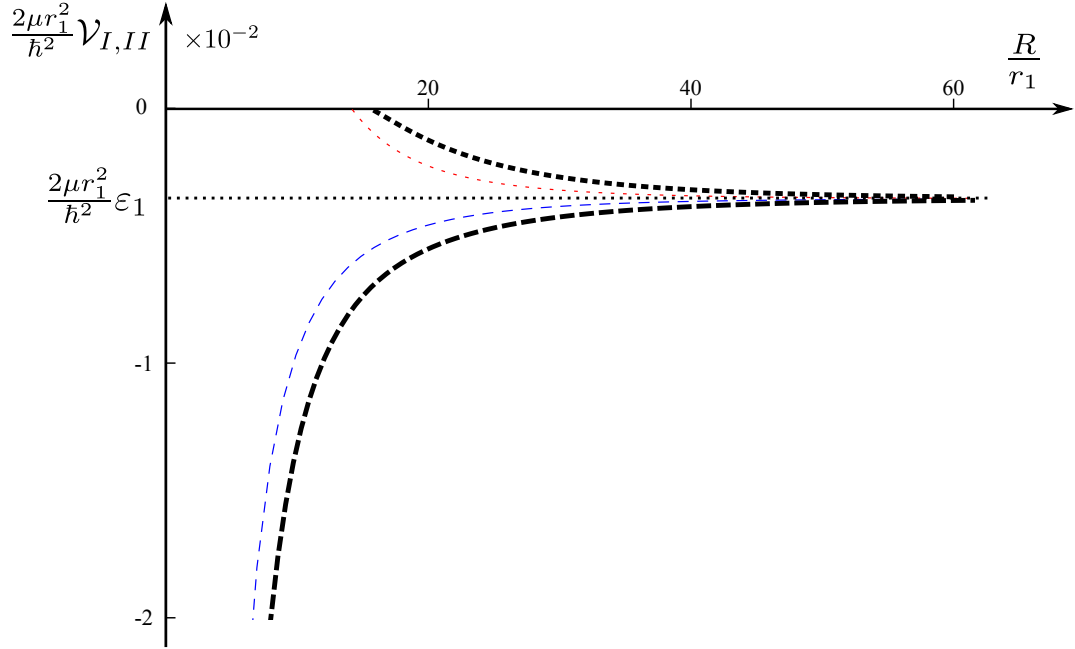}\\
  \caption{Effective potentials ${\mathcal V}_{\rm II}^{(+)}={\mathcal V}_{\rm II}^{(+)}(R)$ (dashed black curve)
  and ${\mathcal V}_{\rm II}^{(-)}={\mathcal V}_{\rm II}^{(-)}(R)$ (dotted black curve)
  induced by a $p$-wave resonant two-body interaction in two space dimensions and corresponding to the second branch, equation (\ref{energy-asymptotic-pm-second}),
  compared and contrasted to the effective potentials ${\mathcal V}_{\rm I}^{(+)}={\mathcal V}_{\rm I}^{(+)}(R)$ (dashed blue curve)
  and ${\mathcal V}_{\rm I}^{(-)}={\mathcal V}_{\rm I}^{(-)}(R)$ (dotted red curve) of the first branch,
  equations (\ref{energy-asymptotic-first-definition}) and (\ref{energy-asymptotic-pm-first}).
  Here $\varepsilon_1$ is the energy of the $p$-wave bound state
  given by equations (\ref{energy-p-wave}) and (\ref{energy-p-wave-defnition})
  for the $p$-wave scattering length $a_1=100\,r_1^2$ and the $s$-wave scattering length $a_0=10\,r_1$, with $r_1$ being the $p$-wave effective range.
  Whereas for short distances there is a small deviation between the potentials of different branches, asymptotically,
  that is for $R\rightarrow \infty$, they are identical.}
\end{figure}

In the neighborhood of the resonance, $a_1\gg r_1^2$, the effective potentials
\begin{equation}
 \label{energy-asymptotic-pm-second}
    {\mathcal V}_{{\rm II}}^{(\pm)}\equiv -\frac{\hbar^2}{2\mu r_1^2}\left(\xi_{{\rm II}}^{(\pm)}\right)^2
\end{equation}
are determined by the two solutions $\xi_{{\rm II}}^{(\pm)}$ of equation (\ref{p-wave condition 2-modified}) and presented in Fig. 4.

\section{Quasi-Coulomb energy spectrum}

We are now in the position to address the two-dimensional dynamics of the two heavy bosonic particles dictated by the
Schr\"odinger equation (\ref{Schrodinger heavy}) with the potential ${\mathcal V}$ given by equation (\ref{V asymptotic}), and
induced by the $p$-wave resonance in the heavy-light interaction.
In particular, we show that ${\mathcal V}$ supports an infinite number of three-body bound states following a quasi-Coulomb series.
Here we first focus our discussion on an exact resonance and then relax this requirement.

For the case of a vanishing orbital angular momentum between the two heavy bosonic particles,
equation (\ref{Schrodinger heavy}) describing their relative motion reduces to the radial equation
\begin{equation}
 \label{Schrodinger heavy radial}
    \left\{\frac{d^2}{dR^2}+\frac{1}{R}\frac{d}{dR}+\frac{M}{\hbar^2}\left[E-{\mathcal V}(R)\right]\right\}\chi_E(R)=0
\end{equation}
with the WKB-solution
\begin{equation}
 \label{WKB solution}
    \chi_E(R)\cong\frac{{\mathcal N}_E}{\left\{(MR^2/\hbar^2)[E-{\mathcal V}(R)]\right\}^{\frac{1}{4}}}\sin\left[\varphi(R,R_E)\right]
\end{equation}
derived in the Appendix. Here ${\mathcal N}_E$ is the normalization constant and the phase
\begin{equation}
 \label{WKB phase}
    \varphi(R,R_E)\equiv\frac{1}{\hbar}\int_{R}^{R_E}dR'\sqrt{M[E-{\mathcal V}(R')]}+\theta_E
\end{equation}
is accumulated between $R$ and the outer turning point $R_E$ determined by the condition ${\mathcal V}(R_E)=E$.
Moreover, we emphasize that the phase $\theta_E$ with $|\theta_E|\leq \pi$,
which depends only weakly on the energy $E$, is determined by the behavior of
the potential at short distances, that is for $R\sim r_1 \sim r_0$.

\subsection{Exact $p$-wave resonance}

Since we are interested in the spectrum of the bound states in ${\mathcal V}$ close to the threshold $E=0$,
we need to know the behavior of $\varphi$ as $E\rightarrow 0$. Indeed,
in the limit of a slightly negative energy, $|E|\ll \hbar^2/(M r_1^2)$, and large distances, $r_1\ll R \ll R_E$, we show in the Appendix that
we can neglect the energy $E$ under the square root on the right-hand side of equation (\ref{WKB phase}) and
obtain with equation (\ref{V asymptotic}) the approximation

\begin{equation}
 \label{WKB phase approximate first}
    \varphi(R,R_E)\cong 2\sqrt{\frac{M}{\mu}}\left[\left(\ln\frac{R_{E}}{r_1}\right)^{\frac{1}{2}}-
    \left(\ln\frac{R}{r_1}\right)^{\frac{1}{2}}\right]+\theta_0
\end{equation}
for the accumulated phase.

The energy spectrum of the weakly bound states caught in the potential ${\mathcal V}$
follows from the familiar WKB quantization rule
\begin{equation}
 \label{WKB-quantization}
    \varphi(r_1,\rho_{n})=\pi n,
\end{equation}
giving rise to the discrete positions
\begin{equation}
 \label{WKB-quantization-Rn}
    \rho_{n}\cong r_1\exp\left(\frac{\pi^2}{4}\frac{\mu}{M}\,n^2\right)
\end{equation}
of the outer turning points for $n\gg 1$.

The connection $E_n={\mathcal V}(\rho_{n})$
between the binding energy $E_n$ and $\rho_{n}$ finally yields with equation (\ref{V asymptotic}) the asymptotic energy spectrum
\begin{equation}
 \label{energy spectrum}
    E_n=-\frac{E_0}{n^2}\exp\left(-\frac{\pi^2}{2}\frac{\mu}{M}\,n^2\right)
\end{equation}
for $n\gg 1$ in the form of the Coulomb series with a Gaussian cut-off governed by small mass ratio $\mu/M$.
The characteristic energy $E_0\sim \hbar^2/(\mu r_0^2)$ is determined by the short-range physics of the system under consideration.

Since all parameters of the short-range interaction are absorbed
into the characteristic energy $E_0$, the spectrum given by equation (\ref{energy spectrum}) has a universal form and is solely determined
by the mass ratio $\mu/M$ as shown in Fig. 1.

\subsection{In the neighborhood of a $p$-wave resonance: number of three-body bound states}

An exact two-dimensional $p$-wave resonance in the heavy-light short-range interaction potential, that is $a_1^{-1}=0$,
creates the two long-range effective potentials ${\mathcal V}_{{\rm I}}^{(0)}$ and
${\mathcal V}_{{\rm II}}^{(0)}$ given by equations (\ref{energy-asymptotic-first}) and (\ref{energy-asymptotic-second}), respectively.
They both merge into the same asymptotic potential
${\mathcal V}$, equation (\ref{V asymptotic}), which gives rise to an infinite series of the weakly bound three-body states.

However, near the resonance, that is for large but finite values of $a_1$,
the range $R_1$ given by equation (\ref{range R_1}) is finite and the effective potentials
${\mathcal V}_{\rm I}^{(0)}$ and ${\mathcal V}_{\rm II}^{(0)}$ are valid only in the region $r_1\lesssim R\lesssim R_1$.
As a result, the number $N_b$ of bound states supported by these potentials is finite and
given by the number of nodes of the zero-energy solution $\chi_0$
defined by equation (\ref{WKB solution}) with the phase $\varphi(R,R_1)$, equation (\ref{WKB phase approximate first}),
accumulated between $R$ and $R_E\sim R_1$,

Since $R_1\gg r_1$, we can estimate the number
$$
N_b\cong \frac{1}{\pi}\,\varphi(r_1,R_1)
$$
of the three-body bound states with the help of equation (\ref{range R_1}) as
\begin{equation}
 \label{N bound states}
    N_b\cong\frac{2}{\pi}\left(\frac{M}{\mu}\ln\frac{R_1}{r_1}\right)^{\frac{1}{2}}\cong
    \frac{1}{\pi}\left(2\,\frac{M}{\mu}\ln\frac{a_1}{2r_1^2}\right)^{\frac{1}{2}}.
\end{equation}

Thus, $N_b$ increases with the square root of $M/\mu$ and diverges weakly as a logarithm of $a_1/r_1^{2}$
when we tune $a_1$ closer to the $p$-wave resonance, that is for $a_1\rightarrow \infty$.
For this reason we obtain an infinite number of three-body bound states in the limit of an exact $p$-wave resonance.

\section{Resonances in atom-molecule scattering confirm quasi-Coulomb series}

We can verify experimentally the existence of the binding potentials ${\mathcal V}_{\rm I}^{(0)}$ and ${\mathcal V}_{\rm II}^{(0)}$
given by equations (\ref{energy-asymptotic-first}) and (\ref{energy-asymptotic-second}), respectively,
by scattering \cite{atomic-molecule, Hammer-Petrov} a heavy atom off the diatomic molecule consisting of the heavy and the light atom.
The predicted three-body bound states manifest themselves as resonances in the cross section of the
atom-molecule scattering when we tune the scattering length using Feshbach resonances \cite{Feshbach-review} and
approach the two-dimensional $p$-wave resonance.

At the low incident energy $(\hbar k)^2/M\equiv E-\varepsilon_1$, such as
$kr_1\ll 1$, the total atom-molecule cross-section \cite{LL}
\begin{equation}
  \label{atom-molecule-sigma}
    \sigma_0=\frac{\pi^2}{k}\left[\frac{\pi^2}{4}+\ln^2\left(\frac{kA_0}{2}\,e^{\gamma}\right)\right]^{-1}
\end{equation}
is determined mainly by the two-dimensional atom-molecule scattering
length $A_{0}$.

In order to estimate the atom-molecule scattering length $A_{0}$ for large values of $a_1$
we need to solve equation (\ref{Schrodinger heavy radial})
with $E=0$ and the potential ${\mathcal V}$ given by equation (\ref{V asymptotic}).

For the large distances $R\gtrsim R_1$, where the effective potential ${\mathcal V}$ vanishes,
the zero-energy solution $\chi_0(R)$ of the radial Schr\"odinger equation (\ref{Schrodinger heavy radial})
reads $\chi_0(R\gtrsim R_1)\propto\ln(R/A_0)$.
In contrast, for the small distances $R\lesssim R_1$,
the solution $\chi_0(R)$ is derived in Appendix and has the form $\chi_0(R\lesssim R_1)\propto\sin[\varphi(r_1,R)]$
with the phase $\varphi(r_1,R)$ given by equation (\ref{WKB phase approximate first}).
When we match the logarithmic derivatives of these two solutions at a distance $R\sim R_1$, we find the relation
$$
\frac{1}{\ln(R_1/A_0)}=R_1\frac{\chi'_0(R_1)}{\chi_0(R_1)}=\frac{2M}{\mu}
\frac{\cot\left(\pi N_b\right)}{\pi N_b}
$$
with $N_b$ given by equation (\ref{N bound states}).

Hence, the atom-molecule scattering length
\begin{equation}
  \label{atom-molecule A0}
    A_0=\left(\frac{a_1}{2}\ln\frac{a_1}{2r_1^2}\right)^{\frac{1}{2}}
  \exp\left\{-\frac{\mu}{2M}\frac{\pi N_b(a_1)}{\cot\left[\pi N_b(a_1)\right]}\right\}
\end{equation}
exhibits an infinite series of resonances at
$a_1=a_1^{(n)}$, where $a_1^{(n)}$ are the zeros of the equation
\begin{equation}
  \label{atom-molecule-equation}
    N_b\left(a_1^{(n)}\right)=\left(n+\frac{1}{2}\right).
\end{equation}

For $n\gg 1$, equation (\ref{N bound states}) provides us with the expression
\begin{equation}
  \label{atom-molecule a_1^n}
    a_1^{(n)}\sim r_1^2\exp\left(\frac{\pi^2}{2}\frac{\mu}{M}n^2\right)
\end{equation}
for the position of the resonances in the total atom-molecule cross-section $\sigma_0$ given by equation (\ref{atom-molecule-sigma})
as a function of the two-dimensional $p$-wave scattering length $a_1$.
For a large mass-ratio $M/\mu$ the maximum number of observable resonances is given by equation (\ref{n_max}).

\section{Summary and outlook}

We have found a series of bound
states in a three-body system consisting of a light particle and two heavy bosonic ones when the
heavy-light short-range interaction potential has a
two-dimensional $p$-wave resonance, and the system is constrained to two space dimensions.
When the total angular momentum of the three particle vanishes and we tune the heavy-light interaction to an {\it exact} $p$-wave resonance,
the effective potentials between the two heavy particles are attractive and of
long-range. They support an {\it infinite} number of bound states and the corresponding spectrum has the form of the Coulomb series
with a Gaussian cut-off governed by the mass ratio.
We emphasize that these results are a consequence of an intricate interplay between the symmetry properties of
the underlying resonances and the dimensionality of the problem.

We are well-aware of the fact these results are obtained within a Born-Oppenheimer approach and questions about its validity are justified.
In order to answer them, we are presently developing a computer code based on parallel programming \cite{DataVortex},
which solves the Schr\"odinger equation for three interacting particles in two space dimensions with arbitrary masses and
the heavy-light interaction being fine-tuned to the $p$-wave resonance.
This approach is important from both theoretical and experimental points of view, since it will allow us
to compare our approximate but analytical predictions to exact numerical solutions and those based on an adiabatic hyperspherical expansion
\cite{super-Efimov-Gridnev, super-Efimov-Volosniev}.

For any experiment involving mixtures of cold atoms with a large mass ratio
the gravitational sag is an important issue to be reckoned with,
especially in observing the quasi-Coulomb spectrum given by equation (\ref{energy spectrum-intoduction}).
Indeed, different species experience different gravitational potentials.
This fact leads to a shift in the centre of the trap, and for a large mass ratio it could even
result in a complete spatial separation of the individual atomic gases.

This effect can be reduced in the laboratory by a rather sophisticated arrangement of bichromatic trapping potential \cite{Efimov-Heidelberg}
created by overlapping optical dipole traps with two distinct wavelengths.
Another very timely approach consists of bringing the experiment into a microgravity environment such as a drop tower \cite{drop-tower-Bremen},
an airplane in parabolic flight \cite{parabolic flight}, or a sounding rocket \cite{sounding rocket} with
the International Space Station \cite{ISS} being ultimate long-time low gravity platform.
We are confident that such extreme conditions will lead to the experimental confirmation of the predicted quasi-Coulomb spectrum.

\ack
We are deeply indebted to S. Batelu, K. Bongs, E.A. Cornell, L. Happ, J.P. D'Incao,
R. D\"orner, P.W. Engels, F. Ferlaino, C.H. Greene, R. Kaiser, S. Moroz,
T. Schmid, T. Pfau, E.M. Rasel, J. Ulmanis, and M. Zimmermann for stimulating discussions.
We are very grateful to C. Devany and C.S. Reed for allowing us access to the fast-performing computers of Data Vortex Technology.

This work was supported by a grant from the Ministry of Science, Research and the Arts of
Baden-W\"urttemberg (Az: 33-7533-30-10/19/2) and
the German Space Agency (DLR) with funds provided by the Federal Ministry for Economic Affairs and Energy
(BMWi) due to an enactment of the German Bundestag under Grants No. DLR 50WM1152-1157 (QUANTUS-IV) and
the Centre for Quantum Engineering and Space-Time Research QUEST.
We also appreciate the funding by the German Research Foundation (DFG) in the framework of the SFB/TRR-21.

W.P.S. is grateful to Texas A$\&$M University for a Texas A$\&$M University Institute for Advanced Study (TIAS) Faculty Fellowship.
M.A.E. thanks the Alexander von Humboldt-Stiftung and the Russian Foundation for Basic Research (10-02-00914-a).

\begin{appendix}

\section{Relative motion of the heavy particles: WKB approach}

In this Appendix we derive the solutions of equation (\ref{Schrodinger heavy radial}) for two different energies:
(i) an exact solution for $E=0$, and (ii) an approximate solution for a slightly negative energy $E$.
In the latter case we take advantage of the WKB approach \cite{LL}.

\subsection{Exact solution for zero energy}

For this purpose we first apply the Langer transformation \cite{Langer}
\begin{equation}
 \label{new variable Langer}
  R\equiv r_1e^{x}
\end{equation}
of the coordinate $R$ with $x\geq 0$ and cast equation (\ref{Schrodinger heavy radial}) into the form
\begin{equation}
 \label{Schr-Langer}
    \frac{d^2}{dx^2}\,\chi+\left(\varepsilon e^{2x}+\frac{\nu_0}{x}\right)\chi=0.
\end{equation}
Here we have introduced the dimensionless energy
$$
\varepsilon\equiv \frac{Mr_1^2 E}{\hbar^2}
$$
and the effective strength
$$
\nu_0\equiv \frac{M}{\mu}
$$
of the one-dimensional Coulomb potential.

For the total three-body energy $E=0$, that is $\varepsilon=0$, equation (\ref{Schr-Langer})
coincides with the Schr\"odinger equation for the one-dimensional Coulomb potential \cite{Loudon} $-\nu_0/x$ at zero energy,
which with the transformation $4\nu_0 x\equiv\zeta^2$ turns into
\begin{equation}
 \label{Schr-modified}
    \left[\frac{d^2}{d\zeta^2}+\frac{1}{\zeta}\frac{d}{d\zeta}+\left(1-\frac{1}{\zeta^2}\right)\right]\frac{\chi_0}{\zeta}=0.
\end{equation}

When we recall the differential equation \cite{Abramowitz}
$$
\frac{d^2}{d\zeta^2}Z_n+\frac{1}{\zeta}\frac{d}{d\zeta}Z_n+\left(1-\frac{n^2}{\zeta^2}\right)Z_n=0
$$
for the Bessel functions $Z_n=Z_n(\zeta)$, we find the analytical solution
\begin{equation}
 \label{exact solution}
    \chi_{0}(x)=\sqrt{x}\left[AJ_1\left(2\sqrt{\nu_0 x}\right)+BY_1\left(2\sqrt{\nu_0 x}\right)\right]
\end{equation}
in the form of a linear superposition of the Bessel functions $J_1$ and $Y_1$ \cite{Abramowitz} with coefficients $A$ and $B$.

Since the potential ${\mathcal V}$ given by equation (\ref{V asymptotic}) is valid only for $R\gg r_1$, that is $x\gg 1$,
we can use the asymptotic expansion of the Bessel functions for large arguments.
As a result, the exact zero-energy solution $\chi_0$, equation (\ref{exact solution}), reads
\begin{equation}
 \label{exact solution-large distance}
    \chi_{0}(x)\propto x^{\frac{1}{4}}\sin\left(2\sqrt{\nu_0 x}-\theta_0\right),
\end{equation}
where the phase
\begin{equation}
 \label{exact solution-large distance-theta0}
    \theta_0\equiv \arctan\left(\frac{B}{A}\right)+\frac{\pi}{4}
\end{equation}
is determined by the potential at short distances $x\sim 1$.

\subsection{Approximate solution for a slightly negative energy}

Next we obtain the solution of equation (\ref{Schr-Langer}) in the case of a slightly negative energy $\varepsilon$ by an analytical
continuation of the zero-energy solution $\chi_0$ given by equation (\ref{exact solution}).
Here we take advantage of the familiar WKB approach and find the approximate solution
in the domain $1\ll x \ll x_\varepsilon$ governed by the outer turning point $x_\varepsilon$ following from the condition
$x_\varepsilon e^{2x_\varepsilon}=\nu_0/|\varepsilon|$.

Indeed, the WKB solution
\begin{equation}
 \label{WKB solution-appendix}
    \chi_{\varepsilon}(x)\cong {\mathcal N}_{\varepsilon}\left(\frac{x}{\nu_0-|\varepsilon|x e^{2x}}\right)^{\frac{1}{4}}
    \sin\left[\varphi(x,x_{\varepsilon})\right]
\end{equation}
with the quasi-classical phase
\begin{equation}
 \label{WKB phase -appendix}
    \varphi(x,x_\varepsilon)\equiv\sqrt{\nu_0}\int_{x}^{x_\varepsilon}\frac{dx'}{\sqrt{x'}}
    \left(1-\frac{|\varepsilon|}{\nu_0}x' e^{2x'}\right)^{\frac{1}{2}}+\theta_\varepsilon
\end{equation}
reproduces the exact solution $\chi_0$ given by equation (\ref{exact solution-large distance}) in the limit $\varepsilon\rightarrow 0$.
Here ${\mathcal N}_{\varepsilon}$ is the normalization constant and the phase $\theta_\varepsilon$, which is weakly dependent on
the energy $\varepsilon$ and is determined by the potential at short distances $x\sim 1$.

In order to find the energy spectrum we need to know the behavior of the phase
$\varphi$ defined by equation (\ref{WKB phase -appendix}) for
small negative energies, $|\varepsilon|\ll 1$, and at large distances.
For this purpose, we cast equation (\ref{WKB phase -appendix}) into the form
\begin{equation}
 \label{WKB phase-first}
   \varphi(x,x_\varepsilon)=\sqrt{\nu_0}\int_{x}^{x_\varepsilon}\frac{dx'}{\sqrt{x'}}+\theta_\varepsilon-\Phi
\end{equation}
and show that the correction
\begin{equation}
 \label{WKB phase -Phi}
   \Phi\equiv\sqrt{\nu_0}\int_{x}^{x_\varepsilon}\frac{dx'}{\sqrt{x'}}
    \left[1-\left(1-\frac{|\varepsilon|}{\nu_0}x' e^{2x'}\right)^{\frac{1}{2}}\right]
\end{equation}
decreases as $x_\varepsilon^{-\frac{1}{2}}$ as $x_\varepsilon\rightarrow \infty$.

Indeed, when we make in equation (\ref{WKB phase -Phi}) the change of variable, $x'\equiv x_\varepsilon-\xi$, we arrive at
\begin{equation}
 \label{WKB phase-integral}
   \Phi=\sqrt{\frac{\nu_0}{x_\varepsilon}}\int_{0}^{x_\varepsilon-x}
   \frac{\sqrt{1-\xi/x_\varepsilon}}{1+\sqrt{1-(1-\xi/x_\varepsilon)e^{-2\xi}}}\,e^{-2\xi}\,d\xi.
\end{equation}

In the case of $x_\varepsilon\rightarrow \infty$ we can simplify the integral on the right-hand side of equation (\ref{WKB phase-integral})
to give rise to
\begin{equation}
 \label{WKB phase-integral-result}
  \Phi\cong\sqrt{\frac{\nu_0}{x_\varepsilon}}\int_{0}^{\infty}
   \frac{e^{-2\xi}}{1+\sqrt{1-e^{-2\xi}}}\,d\xi=(1-\ln 2)\sqrt{\frac{\nu_0}{x_\varepsilon}},
\end{equation}
where we have used the transformation $1-e^{-2\xi}\equiv z^2$.

Finally we use equations (\ref{new variable Langer}), (\ref{WKB phase-first}), and (\ref{WKB phase-integral-result}) and obtain
the behavior
\begin{equation}
   \varphi(R,R_E)\cong 2\sqrt{\frac{M}{\mu}}\left[\left(\ln\frac{R_E}{r_1}\right)^{\frac{1}{2}}-
    \left(\ln\frac{R}{r_1}\right)^{\frac{1}{2}}\right]+\theta_E
\end{equation}
of the WKB-phase for a slightly negative energy $E$ and large distances, that is $r_1\ll R \ll R_E$.

\end{appendix}

\end{document}